\documentclass[12pt,amsfonts]{article}
\usepackage{graphicx}
\usepackage{latexsym}
\usepackage{amssymb}
\usepackage{amsmath}
\usepackage{enumerate}
\usepackage{dsfont}
\usepackage{xcolor}
\usepackage[urlcolor=black, colorlinks=true, linkcolor=blue, citecolor=brown]{hyperref}

\usepackage{layout}
\usepackage{eufrak}

\makeindex

\makeatletter

\@addtoreset{equation}{section}
\makeatother

\begin{document}

\begingroup

\title{Modeling First Line Of An Order Book With\\ Multivariate Marked Point Processes }
\author{ Alexis Fauth $^{1, 2}$ \qquad %
Ciprian A. Tudor $^{3,4}$ \footnote{Supported by the CNCS grant PN-II-ID-PCCE-2011-2-0015. Associate member of the team Samm, Universit\'e Panth\'eon-Sorbonne Paris 1 } \\
\small $^{1}$SAMM, Universit\'e Paris 1 Panth\'eon-Sorbonne, Paris, France. \\
\small $^{2}$Invivoo, Courbevoie, France. \\
\small  alexis.fauth@invivoo.com\\
\small  $^{3}$ Laboratoire Paul Painlev\'e, Universit\'e de Lille 1, Villeneuve d'Ascq, France.\\
\small  $^{4}$ Department of Mathematics, Academy of Economical Studies, Bucharest, Romania.\\
\small  tudor@math.univ-lille1.fr}

\maketitle

\begin{abstract}
We introduce a new model in order to describe the fluctuation of tick-by-tick financial time series. Our model, based on marked point process, allows us to incorporate in a unique process the duration of the transaction and the  corresponding volume of orders.

The  model is motivated by the fact that the "excitation" of the market is different in periods of time with low exchanged volume and high volume exchanged. We illustrate our result by numerical simulations on foreign exchange data sampling in millisecond. By checking the main stylized facts, we show that the model is consistent with the empirical data. We also find an interesting relation between the distribution of the volume of limited order and the volume of market orders. To conclude, we propose an application to risk management and we introduce a forecast procedure.
\end{abstract}


\section{Introduction}

Buy and sell orders did not arrive at continuous times, counterparts do not meet them at any time, hence, the fluctuations of the stock market cannot evolve continuously and high frequency financial data have a flagrant discontinuity. The idea to use   point processes  to describe such irregularities appears as an  evidence.

The family of stochastic processes that we will use in our analysis, the so-called Hawkes processes, was introduced in the seventies by A. Hawkes in the paper \cite{Ha71a}. The initial purpose was to model earthquake occurrences. A Hawkes process  is a generalization of a standard point process. They already found  applications in various fields. We refer, among others, to Ogata \cite{Og98, Og06} for statistical earthquake modeling, to  Chavez-Demoulin et al. \cite{ChHa09} and  Errais et al. \cite{Er10} for risk analysis, A\"\i t-Sahalia et al. \cite{Aï11} for the so-called financial contagion, to Bacry et al. \cite{Ba11a} for high-frequency financial data and to  Muni Tokes \cite{To10} for order book modeling.\\

The scientific literature related to high frequency data analysis usually proposes to use self-exciting point  processes that naturally correspond to the frequency of the arrival times of orders and do not take into account the corresponding volumes  or the current return. Generally,  in these models,  at the  same frequency, 'small' or 'large' volumes traded will impact the market in the same way.

The literature on stylized facts for high frequency data is very rich. We recall the following properties frequently observed on the data:  fat tails of financial returns, long memory on square returns, the multifractal nature, etc. For  extensive reviews, see e.g. Dacorogna et al. \cite{Da01}, Engle \cite{En00}, Bouchaud and Potters \cite{BoPo04} or Voit \cite{Vo05}. In the case of the very high frequency, which means transactions recorded tick by tick, the most common properties are the so-called Epps effect (see Epps \cite{Ep79}), the signature plot (see Andersen et al. \cite{An00}), the intertrade duration (see Hautsch, \cite{Ha12}), and the market impact (see Moro et al. \cite{Mo09}).\\

On the market, the price changes "tick by tick". The tick size is the minimum of the variation of the price. The price of an asset is written as $p=np_{\text{tick}}$ where $n\in\mathbb{N}^{*}$ and $p_{\text{tick}}$ is the tick size. For example, it is between $10^{-4}$ and $0.5\times 10^{-5}\$$ for the Forex market and usually  0.25$\$$ for future contracts. There are two principal types of orders which can move the price: the market order and the limited order.

A market order will immediately be executed at the best available  price and at the relevant time. Of course, a buy market order will be  executed to the best ask price, while sell market order will be executed to the best bid price. The best ask (bid) price is understood at the lowest price at which an agent can buy (sell) an asset. The volume offered corresponding to best ask price need to be completely bought to see the best ask (bid) price move upward (downard) from one tick.

A limited order is an order to buy (sell) an asset to a specific price, lower (higher) than the best ask (bid) price. Then, a new offered volume inside the bid-ask spread will move the price downward or upward depending if it is proposed on ask side or bid side.

A market order will decrease the amount of shares available on the market, the volume, while a limited order inside the bid-ask spread will increase them. We schematize these two types of orders that may be vary the price in the figure (\ref{OrderBook}). Thus, it is the exchange volume which affects the price and the financial fluctuation. During the excitation phase, i.e. when the price moves quickly, the market volume is quickly exchanged and hence, the prices vary rapidly. The time transaction and the volume are strongly dependent since one produces the other one. Several papers related to the theory of  market impact try to deals with this aspect, see e.g. E. Moro et al. \cite{Mo09}, or Z. Eisler et al. \cite{EiBoKo10}. Our purpose is to incorporate this information concerning the variable "volume of orders" in our model. To do this, we will employ the marked Hawkes process, their mark corresponding to the volume of each transaction.\\

\begin{figure}[h!]
      \centering \includegraphics[height=5cm, width=14cm]{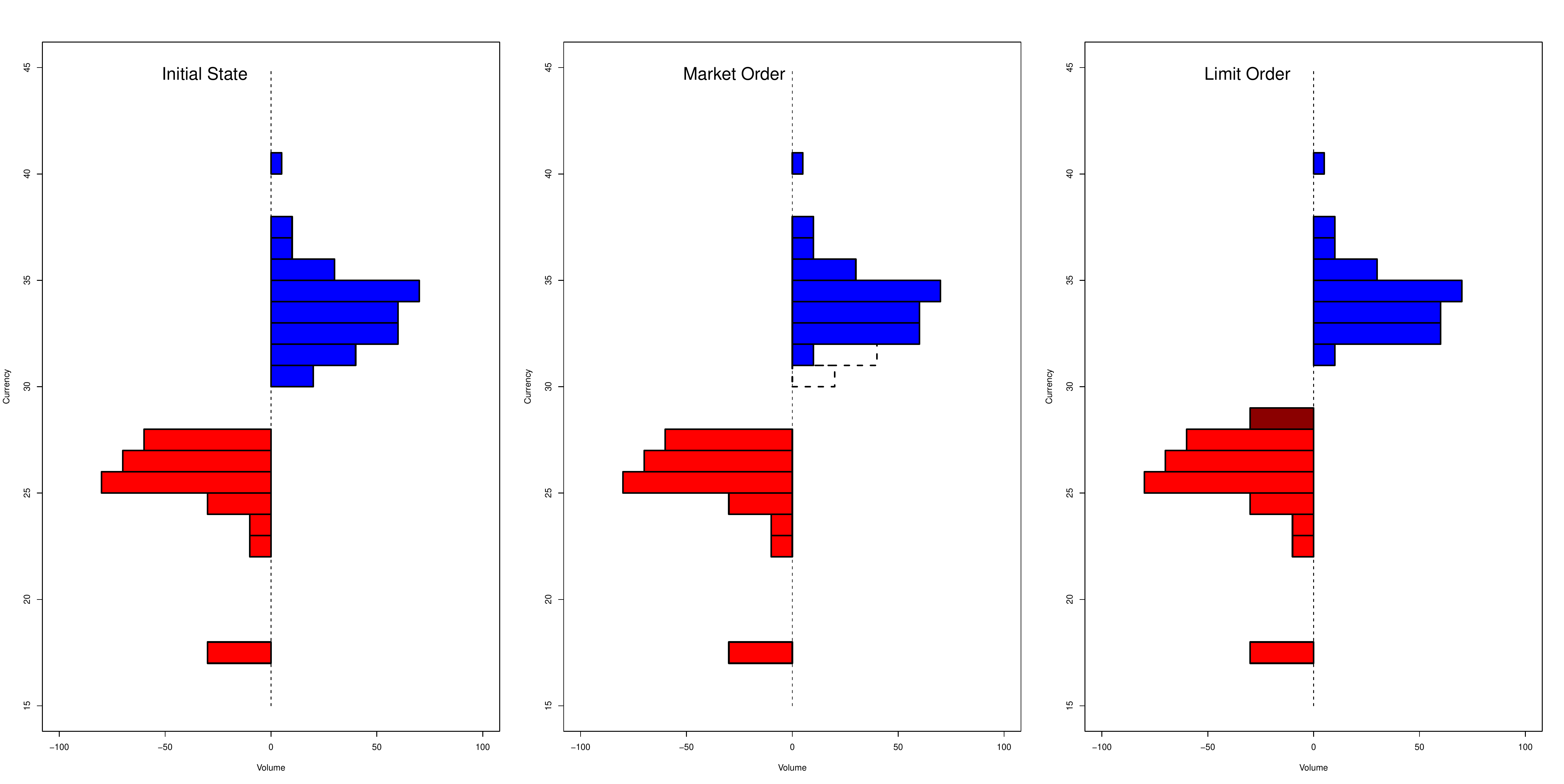}
\caption{Schematic representation of an Order Book. In red volume proposed for sell, in blue volume proposed for buy. Left: initial state. Middle: same order book after a buy market order which move upward the best ask form one tick. Right: a new buy limited order inside the bid-ask spread which move upward the best bid.}
\label{OrderBook}
\end{figure}

In order to  reproduce the fluctuation of the price, we have to take into account the mechanism of the formation of the price. Consider for example a limited order book. Hence, we can observe the market order and the limited order.  We have conscience of the existence of canceled orders too, but, since we do not have this information only looking to the order book, we cannot take it  into account  in our framework. By analyzing  some SEC (Securities and Exchange Commission) reports or their counterparts in France, the AMF (Autorit\'e des March\'es Financiers) reports, we can notice that  about 80$\%$ of orders are canceled. As this amount is not tangible, except by the regulatory authorities, we choose to avoid that quantity. Then, the appearance of these different types  of orders (marker orders and limited orders) produces the price. If we are able to reproduce this scheme, we should be able to reconstruct the fluctuation of the price.\\

The aim of this study is to build a model with  two approaches, take into account in a certain way the limited order book but focusing on  the best bid and the  best ask prices. The mathematical tool used is represented by the marked Hawkes process. To test how realistic is  our model, we will check the different stylized facts. In particular, to reproduce the Epps effect, that is,  the correlation between two assets at different frequencies, the model will be presented in a multivariate form.\\

First, we recall some existing models related to our problem and we will explain why we choose  our approach. After, we  recall the definition and the basic properties of the multivariate marked Hawkes processes and using them,  we will introduce our model in $d$ dimensions. We conclude by comparing the empirical data and the stylized facts  to our model and we give some ideas on how  to apply this result to trading and risk management. All numerical  computations  are done on foreign exchange market.

\section{Literature Review}

We have a very interesting literature on modeling very high frequency data using point processes, and in particular  Hawkes processes. Basically, we have two main approaches.

 Let us recall the main ideas of these two approaches. In the first  one, one   tries to mode the data using only the tick information, see e.g. \cite{Ba11a}, which is the frequency of order arrival. Such models have the following form

\begin{equation}
p(t)=N^+(t)-N^-(t),
\end{equation}
where $p(t)$ is the asset price and $N^\pm(t)$ is a counting process corresponding to the number of upward and downward jumps. The "asset price" is considered as the last price, or the mid-price. The mid-price is the mean between ask and bid price, $(p_{ask}+p_{bid})/2$. Unfortunately, in that case, we do not have any information about the bid and ask prices. The spread, which is the difference between bid price and ask price is a well known cost by practitioner. The best bid price is the highest price at which an impatient trader sell his asset, in other words it is  the market price to sell-short an asset. In a similar way, the best ask price is the lowest price at which an impatient trader can buy an asset. Hence, if the agent buys  at market $k p_{ask}=x\$$ $k$ shares of an asset,  where $k$ is the amount of contracts bought  and then needs to sell just after, the agent   has to sell at bid price which is, assuming no fluctuation happened during that laps time, $k p_{bid}=(x\$-s)k$ where $s$ is the  bid-ask  spread,
$$
s=p_{ask}-p_{bid}.
$$
Consequently, the transaction cost is $k s\$$. In order to optimize his trading activity, any agent has to take into account this quantity; of course, we have  to add other inherent costs such as broker fees  and the market impact. Recall that market impact is the effect that buying or selling moves the price against the trader: upward when buying, downward when selling.\\

The second approach used in the literature  is to try to model the  limited order book, which is somehow more  complicated. These models, to be accurate and consistent to the problem, requires to take into account other aspects  and not only the  transaction price and the trade duration. In \cite{La}, J. Large try to work with ten types of orders:
\begin{enumerate}
\item Market buy that moves the ask price
\item Market sale that moves the bid price
\item Bid between the quotes
\item Ask between the quotes
\item Market buy that doesn't move the ask price
\item Market sale that doesn't move the bid price
\item Bid at or below bid price
\item Ask at or above ask price
\item Cancelled bids
\item Cancelled asks
\end{enumerate}
Notice that the aim of the above reference  is to measure the resiliency of an limited order book. Resiliency is the ability to a market to come back to his equilibrium after a large aggressive order. Any of this type of order is modelized by a counting process $N_i$ with intensity $\lambda_i$, $i=1,\cdots,10$.\\

I. M. Toke, \cite{To10} simplifies the formulation by  proposing  to describe a limited order book by a two-agents model. One patient trader who submits only limits orders who sometimes cancels his own orders. The  second agent  is impatient and submits only market orders. In both cases, what is really interesting, is that we have a corresponding volume of orders drawn from an exponential law. The  cancelation  of an order  arrives with a certain probability $\delta$. Other quantities are drawn from a Hawkes process.\\

In both cases, it become complicated and time consuming  to extend such  models in a 2-dimension way and to include  the lead-lag effect or the Epps effect.

\section{Multivariate Marked Hawkes Process}

Let $N(t)$ a $d$-dimensional point process, $N=(N_1,N_2,\cdots,N_d)$, with $N_i(t_i)$, $1\leq i \leq d$ the cumulative number of events for the $i^{th}$ component a time $t_i$. Note that this process $N_{i}(t_{i})$, $i=1,..,d$ must be non-decreasing and right continuous,
and take only non-negative integer values.

All along this paper, we will work on a standard  probability space $(\Omega, \mathcal{F}, \mathbb{P})$ and on this space we will consider $N(t)$ a $d$-dimensional  $\mathcal{F}_t$-adapted process, $t=(t_1,\cdots,t_d)$. To simplify notations, we will often use the notation $t$ instead of  $t_i$ even for point in $\mathbb{R}$.

An important characteristic of the process $N$ is its   conditional intensity  with respect to the filtration considered and this conditional intensity is given by

\begin{equation}
\lambda(t|\mathcal{F}_t)=\lim_{\delta t\searrow0}\frac{1}{\delta t}\mathbb{E}\left[N(t+\delta t)-N(t)|\mathcal{F}_t\right].
\end{equation}
The $\mathcal{F}_t$-intensity characterizes the evolution of the process $N(t)$ with respect to this past history $\mathcal{F}_t$. We could interpret that as the conditional probability at time $t$ to observe a new event at next time $t+\delta t$.

The Hawkes process is a particular class of the point processes. The first definition of the Hawkes process, which is quite specific, has been given in \cite{Ha71a}. Today, one calls a larger class of  point processes as Hawkes processes. This class is characterized by the intensity function. More precisely, the intensity of  of multivariate $d$-dimensional Hawkes process is usually defined by

\begin{equation}
\lambda_i(t|\mathcal{F}_t)=\mu_i+\sum_{j=1}^d\nu_{ij}\int_{-\infty}^th_{i}(t-s)N_j(\mathrm{d}s) \label{inten},
\end{equation}
where $\boldsymbol{h}=\{h_i\}_{i=1,\cdots,d}$ is the so-called decay kernel and satisfies $h_i(t)\ge0$ for all $i$. The parameters $\{\nu_{ij}\}_{i,j=1,\cdots,d}$ are referred as the branching coefficient, it quantifies the ability of an event $i$ to trigger an event of type $j$. Thus, we see the mutually exciting structure of the process since event $j\neq i$ can affect the conditional intensity $\lambda_i$. Self-exciting part is of course the case $j=i$, past and current event will induce a response in their own intensity process and therefore, on the corresponding point process.  The constants $\{\mu_i\}_{i=1,\cdots,d}$ are understood as the rate of instantaneous events.

The branching matrix is the average number of event $j$ triggered by an event of type $i$. Consider the case $d=2$ and the associated branching matrix

\begin{equation}
\boldsymbol{\nu}=\left(\begin{array}{cc} \nu_{11} & \nu_{12} \\ \nu_{21} & \nu_{22}\end{array}\right).
\end{equation}
 In this case $\nu_{12}$ is interpreted as the average number of event of type $i=1$ with parent event type $j=2$. The integral (\ref{inten}) of the kernel with respect to the counting process $N$ can be expressed  as
\begin{equation}
\int_{-\infty}^th_i(t-s)N(\mathrm{d}s)=\sum_{k|t<t_k}h_i(t-t_k),
\end{equation}
where $t$ is the physical time  and $\{t_k\}_{k=1,\cdots,n}$ represents the arrival times of events. If a new event occurs the intensity decreases. This phenomena is described by the decay kernel $\boldsymbol{h}$. A natural and standard choice of the function $\boldsymbol{h}$ is,

\begin{equation}
h_i(t-t_k)=\alpha_i\exp(-\alpha_i(t-t_k)),\: \alpha_i\ge0
\end{equation}
for every $i=1,..,d$. If we observe the arrival of a new event, then the decay function will be close to $\alpha_{i}\ge0$ and if  no new event occurs, then the decay  function will tend towards 0 exponentially.\\

The purpose of this work is to incorporate the volume of the orders in our model. As mentioned before, we will use multivariate Hawkes process but we will  incorporate \color{black} the marks in the intensity expression (\ref{inten}). A mark is an additional value attached to each point and brings some new information about the points. Consequently,  we will have marked intensities, denoted in the sequel $\lambda(t,v_t|\mathcal{F}_t)$ where $v_t$ represents the mark  and will model the volume of the orders at time $t$. The marked intensity (see e.g. Daley and D. Vere-Jones 03, \cite{DaVe03} and Liniger 09, \cite{Li09} for an extensive review) takes the form,

\begin{equation}
\lambda_i(t, v|\mathcal{F}_t)=\mu_i+\sum_{j=1}^d\nu_{ij}\int_{(-\infty,t)\times\mathbb{R}^+}h_{i}(t-s)g_j(v)N_j(\mathrm{d}s\times \mathrm{d}v). \label{mhawkes}
\end{equation}
Now, $\mathcal{F}_t$ is the history of arrival time and corresponding mark, $\{t_i,v(t_i)\}$. The function $g_j$, $j=1,\cdots,d$ is the so-called impact function of marks, in other words, it characterizes the impact of the volume on the financial asset fluctuation. It has to satisfy the condition  $\int_{\mathbb{R}^d}g(v)f(v)\mathrm{d}v=1$ where $f$ is the density of the vector  $V=(V_1,\cdots,V_d), V_i=(v_{i}(1),\cdots,v_{i}(t),\cdots,v_{i}(T))$. Most common choices for the impact function are,

\begin{equation}
\tilde{g}(x)=x^\alpha ,\:\:\alpha>0,\quad \text{ and } \quad \tilde{g}(x)=\exp(\alpha x), \:\:\alpha>0. \label{impfun}
\end{equation}
Naturally, the first one corresponds to an impact in power law, and the  second one means an exponential impact of the volume on the asset price. We used the notation $\tilde{g}$ because we need to normalize  the function in order to respect condition $\int_{\mathbb{R}^d}g(v)f(v)\mathrm{d}v=1$. This normalization condition is mainly made for a comprehensive way than anything else. Using that representation implies that branching coefficient appears explicitly (see \cite{Li09}), and then we could interpret the different relation and deals simply with condition on branching matrix eigenvalues. Then the corresponding normalized versions  of (\ref{impfun}) are:

\begin{equation}
g(x)=\frac{x^\alpha}{\mathbb{E}X},\:\:\alpha>0,\quad \text{ and } \quad g(x)=\frac{\exp(\alpha x)}{\mathbb{E}X}, \:\:\alpha>0
\end{equation}
where $X$ is a random variable with density  $f$.

Without the  impact function, or  in other words without taking into account the  mark, the intensity's increase depends only on the time duration between two events. Using marked Hawkes process, the intensity increases with respect to the arrival time of events but also with respect to the mark value. So intensity, and finally the price, will be influenced by the  trading volume. This is the fundamental reason why we have chosen to work  with this process. Indeed, the market impact depends on the degree of excitation, but not only on the  frequency of the arrival of significant events. It is also affected by the amount of the exchanges  at each moment. It it then  natural  to use marked point process instead of unmarked process.

The kernel $h_{i}(t-s)$, $i=1,2,...,d$  gives us the probability that at time $s<t$, an event of type $j$ to trigger an event $j$ at time $t$. Notice that as for the non marked  and univariate case, we simply have,

\begin{equation}
\lambda_i(t, v|\mathcal{F}_t)=\mu_i+\sum_{j=1}^d\nu_{ij}\sum_{k|t^{(j)}_k<t}h_{i}(t-t^{(j)}_k)g_j(v(t^{(j)}_k)).
\end{equation}

The multivariate marked Hawkes process is well defined and stationary if in (\ref{mhawkes}) we have the spectral radius of the  branching coefficients matrix, denoted $\boldsymbol{\nu}$, less than 1:
\begin{equation}
\max_i|\lambda_i|<1, i=1,\cdots,n
\end{equation}
where $\lambda_1,\cdots,\lambda_n$ are the eigenvalues of the branching matrix. Also, the decay function verifies:
\begin{equation}
\displaystyle\int_0^\infty th_{i}(t)\mathrm{d}t<\infty.
\end{equation}
\color{black}The first condition gives, after numerical parameters estimation, a first response if the Hawkes process defined above fits with the practical situation  or not. Notice that  the coefficients of the matrix $\boldsymbol{\nu}=\{\nu_{ij}\}_{i,j=1,\cdots,d}$ are themselves issued of the estimation. This first response  is a necessary but not a sufficient condition to validate the model. We will discuss further in the section devoted to the estimation procedure. The second condition implies some restriction in the choice of decay kernel function $h$.\\

 Notice that to verify quality simulation, the Hawkes process provides by their nature two goodnest-of-fit tests. First one is that the integrated intensity between two consecutive events follows an exponential law of rate 1,

\begin{equation}
\int_{t_{k-1}}^{t_k}\lambda(t,v|\mathcal{F}_t)\mathrm{d}t\times\mathrm{d}v\sim \text{Exp}(1). \label{fit1}
\end{equation}
The second one is the called random time changed theorem (see Proposition 7.4.VI (b) in \cite{DaVe03}), if $\int_0^\infty\lambda_j(t,v_t|\mathcal{F}_t)=\infty$ for all $j=1,\cdots,d$, then, under the random time transformation,

\begin{equation}
(t,v)\mapsto(\Lambda(t,v),v),\label{fit2}
\end{equation}
the marked point process is transformed into a compound Poisson process $\tilde{N}$ with unit ground rate and stationary mark distribution $f$. \color{black}\\

\section{Modeling high-frequency data}

When someone tries to model an order book, he, she has to be aware that he models only one order book corresponding to one broker. Each broker has his own order book, his own agents market participants. Usually, thanks to arbitragers and market maker, in any case, we have the same last price. That kind of arbitrage opportunity, when the bid and ask prices differ from one broker to another, is quickly detect. On Forex market, due to the multiplication of market places, looking to the action of only  one broker has no really sense, that's why we commonly have just best bid and best ask prices with corresponding volume. That's why we focus only on these two price processes.\\

Let us recall that the two major aims of our study are to give a model able to reproduce financial fluctuations and to take into account the bid-ask spread. To have an idea on the  bid-ask spread, we think that the most simple and realistic model is to diffuse conjointly bid and ask prices, and thus, to deduce the bid-ask spread. We have to take into account the dependence between the two kind of prices and of course, this two quantities are usually  strong correlated.

Also, as we mentioned before, an important volume of trading does not affect market in same manner as a small volume does.
One of  the most popular stylized facts, the so-called cluster of return, implies that  large changes are usually followed by large changes. It is therefore  natural to think that a large exchanged volume could trigger many trades, exciting the market. We have plotted on (\ref{return}) the log-return of EUR/USD in ``deci''-seconds sampling to see these phenomena.

\begin{figure}[h!]
       \centering \includegraphics[height=8cm, width=14cm]{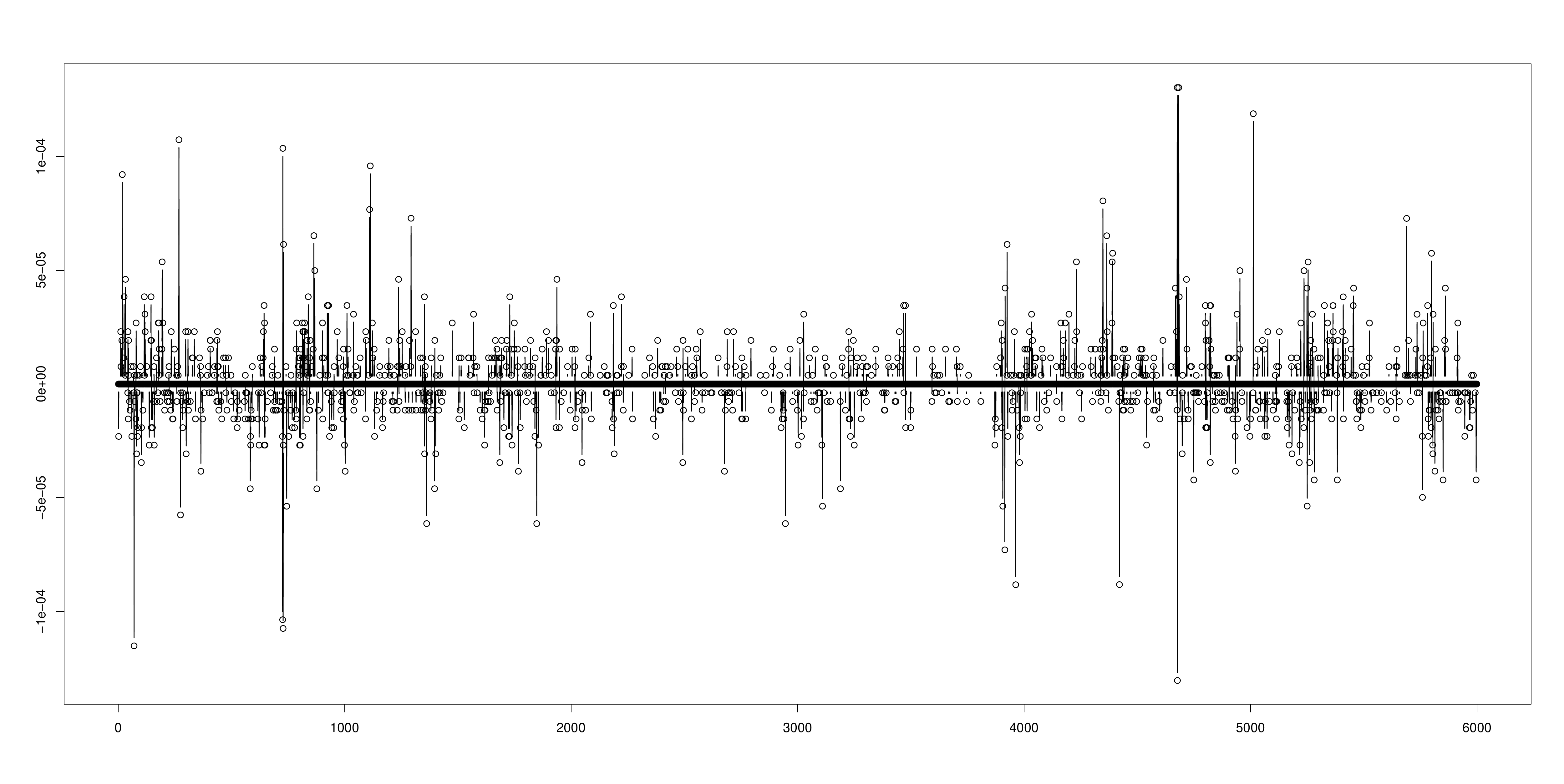}
\caption{EUR/USD log-return from 14:00:00 to 14:10:00, 06.02.2012 in decisecond sampling.}
\label{return}
\end{figure}

To confirm this  idea about the impact of the trading volume on the price we present In Figure (\ref{durvol}) the plot of the time interval between the change of volume and the corresponding volume variation. Data are EUR/USD and EUR/GBP form 30-01-2012 to 10-03-2012 recorded in milliseconds. We clearly see on that plot that a big volume variation corresponds to a short variation. The intertrade time interval (or simply duration), is then defined as the laps time between two consecutive orders,
\begin{equation}
 d_k=t_k-t_{k-1}
\end{equation}

\begin{figure}[h!]
\begin{minipage}[b]{0.5\linewidth}
       \centering \includegraphics[height=7.5cm, width=6.5cm]{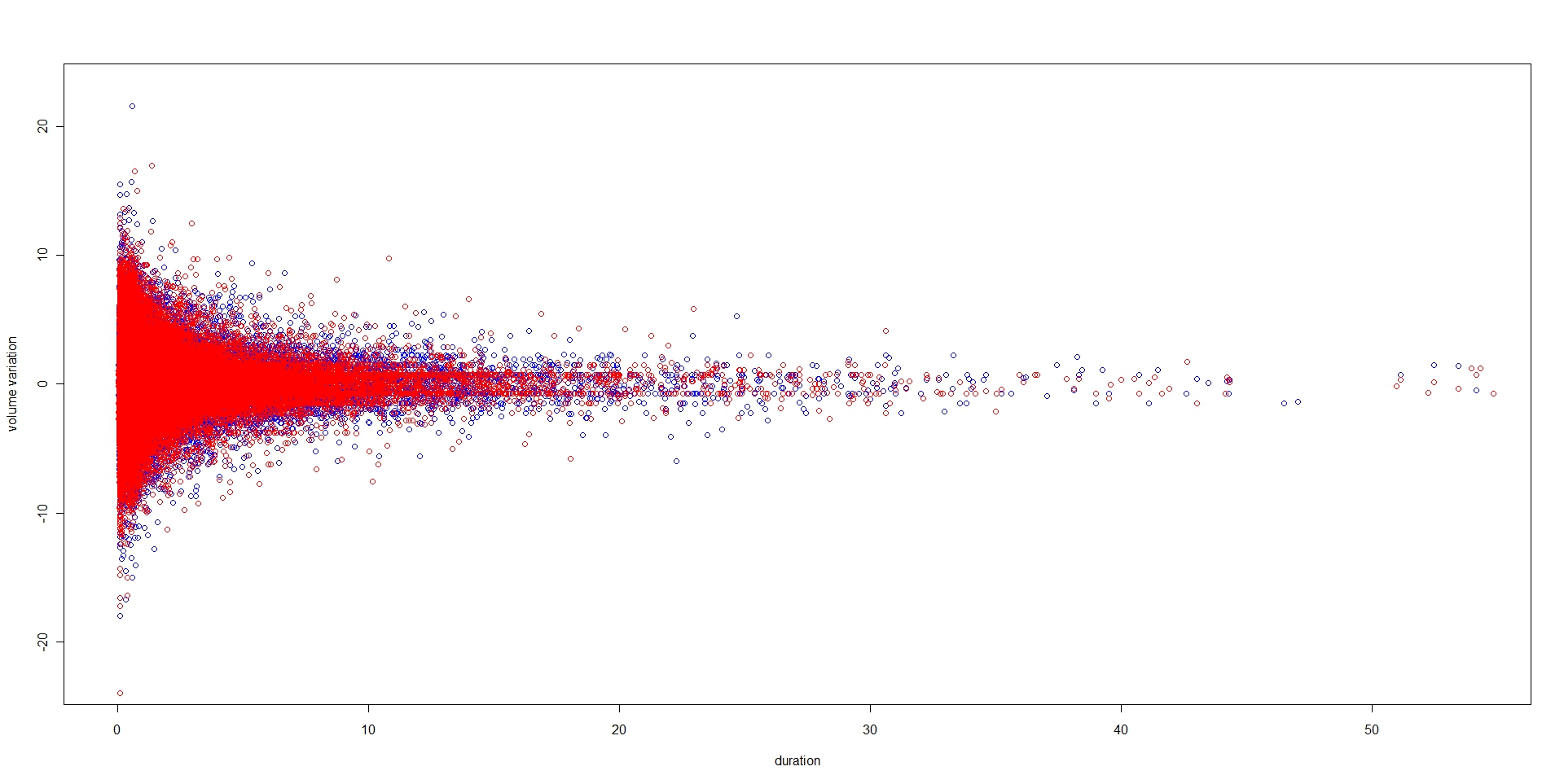}
\end{minipage}\hfill
\begin{minipage}[b]{0.5\linewidth}
       \centering \includegraphics[height=7.5cm, width=6.5cm]{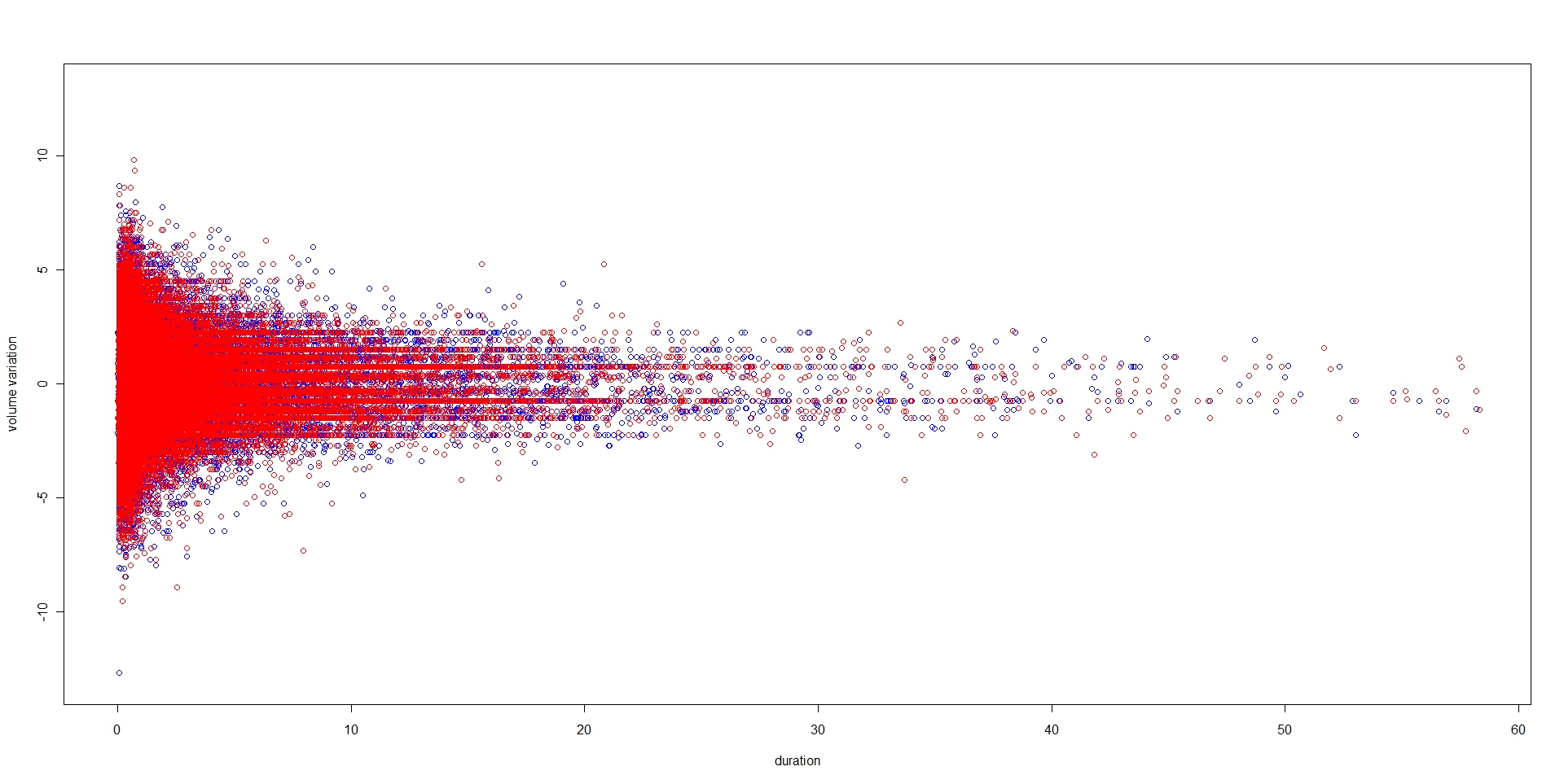}
\end{minipage}
\caption{Volume variation with respect to duration, second scale. In blue (red), the change in ask (bid) volume. Left: EUR/USD, right: EUR/GBP. The variations are given in Million of USD. Notice that the duration near to 1 minute correspond to the price recorded around midnight GMT.}
\label{durvol}
\end{figure}

Let us now explain our model. Writing $a$ for best ask price, $b$ for best bid price, + for upward jump and - for downward jump, our model takes the form:

\begin{equation}
\begin{aligned}
&\text{Ask: }\left\{\begin{array}{ll}
\lambda_{a, +}(t)&=\mu_{a, +}+\displaystyle\sum_{j=a+,b+}\nu_{a,+,j,+}\int_{(-\infty, t)\times \mathbb{R}} h_{a, +}(t-s)g_j(v)N_j(\mathrm{d}s\times\mathrm{d}v)\\
\lambda_{a, -}(t)&=\mu_{a, -}+\displaystyle\sum_{j=a-,b-}\nu_{a,-,j,-}\int_{(-\infty, t)\times \mathbb{R}} h_{a, -}(t-s)g_j(v)N_j(\mathrm{d}s\times\mathrm{d}v)\end{array}\right.\\
&\text{Bid: }\left\{\begin{array}{ll}
\lambda_{b, +}(t)&=\mu_{b, +}+\displaystyle\sum_{j=a+,b+}\nu_{b,+,j,+}\int_{(-\infty, t)\times \mathbb{R}} h_{b, +}(t-s)g_j(v)N_j(\mathrm{d}s\times\mathrm{d}v)\\
\lambda_{b, -}(t)&=\mu_{b, -}+\displaystyle\sum_{j=a-,b-}\nu_{b,-,j,-}\int_{(-\infty, t)\times \mathbb{R}} h_{b, -}(t-s)g_j(v)N_j(\mathrm{d}s\times\mathrm{d}v).\end{array}\right. \label{intenuni}
\end{aligned}
\end{equation}
The above model is univariate, corresponding to only one agent. Let us explain the above notation: the quantity $\lambda _{a,+}$ represents the intensity associated with the ask price when this price had an upward move. Similarly are interpreted the intensities $\lambda_{a,-}, \lambda _{b, +}$ and $\lambda _{b, -}$. The scalar $\nu _{a,+, b, +}$ quantifies  the interaction between the ask price and the bid price (in the upward jump case).

We have made here some choices on the interactions and the dependence between the agents. First, there is no interaction between upward and downward jump, so in the above notation  $\nu_{\cdot,+,\cdot,-}=\nu_{\cdot,-,\cdot,+}=0,$ for all $i,j=a,b$. The dependence between upward jump on bid side and ask side is included in the model and the same holds  for the  downward jump. Let us explain that choices. Since we can naturally think that it  is the bid side (respectively the ask side) which governed the decrease (respectively increase) of the price process, therefore we don't have any process corresponding to the dependence between $+$ and $-$. Since  it is empirically not possible to observe a bid-ask spread tending to infinity we introduced a dependence between upward (downward) jump for the bid and ask prices. Even if they do not evolve in a symmetric way (otherwise the spread will be constant) one follows the other.  The case $j=i$ is then the self-exciting part of the intensity while $j\neq i$ is the mutually exciting part.\\

In this modeling, we deal only with trades that triggers a shift on price. Market order which just pick a part of the offered volume and limited order bigger than the best ask price or lower than the best bid price are not take into account. Indeed, conditional intensity corresponding to best ask price which produce an upward jump correspond to buy market orders, conversely, the best bid price which moves the price downward is a sell market order which completly take the liquidity available on the corresponding tick line. Hence, Since that two quantities attempt to modelise market order, the case of market order who don't move the price is implicitely took into a count by the fact that no event occurs. On the other side, ask (bid) price which trigger an downward (upward) jump is a new limited order inside the bid-ask spread. Then, for the same reasons as before, limited orders outside the bid-ask spread are implicitly took into account when no events occurs.

In one hand, remind that the conditional intensity is the probability to observe a new event knowing the history. On the other hand, like intensities are marked, in other word, they depend not only on the frequency of arrival orders, but also of the corresponding volume. Then, more exhanged volume will be important, more intensity value will be important, and then, more the probability of a new occurence, a new price fluctutation, will increase. Thus, we are consistent with fondamental observation like (\ref{return}) or (\ref{durvol}). \color{black}\\

Since the intensities describe completely the marked point process, knowing $\lambda_j$ permit to simulate a sequence of marked point process $\{t_i,v(t_i)\}_{i=1,\cdots,T}$ (see e.g. \cite{Og81}, or \cite{Li09}) and then, to deduce the counting process $N(t)$. Denoting by $p_a$ and $p_b$ respectively ask and bid prices, we will have

\begin{equation}
\begin{aligned}
p_a(t)&=p(0)+(N_{a,+}(t)-N_{a,-}(t))p_{\text{tick}}\\
p_b(t)&=p(0)+(N_{b,+}(t)-N_{b,-}(t))p_{\text{tick}}
\end{aligned}\label{priceuni}
\end{equation}
where $p_{\text{tick}}$ is the tick size, 10$^{-5}$ for our parities  in our data base, and $N_{a,+}$ constitutes a counting process with intensity $\lambda _{a,+} $ and similarly for $N_{a,-}, N_{b+}, N_{b-}$.\\

We have two distinct formulations for the different counting processes:

\begin{itemize}
\item Consider $N$ as a simple counting process, which is
\begin{equation}
N(t)=\sum_{i=1}^n\mathds{1}_{\{t_i\leq t; v(t_i)=x\}}
\end{equation}

Assuming that at each new event, there is a corresponding mark, indeed, this the case here because a new price fluctuation is necessarly associated to a new quantity of volume added or removed. Then, the counting process take the same value than in a non marked case, except that the frequency will strongly affected by value of volume. Also, using a multivariate point process we have not only a bid and ask price, but also their corresponding volume, $v_{a, \pm}$ and $v_{b,\pm}$.
\item The marked case is an appropriate way to produce compound counting processes. Denote
\begin{equation}
N_g(t)=\sum_{i=1}^{N(t)}v(t_i),\:\: N(t)=\sum_{i=1}^n\mathds{1}_{t_i\leq t}, \label{compound}
\end{equation}
where $g$ stands for  a growth process. Then, if we keep formulation (\ref{compound}) the  price fluctuation would be  directly affected by the volume. This version is not compatible with our representation (\ref{priceuni}) since the price fluctuation is not exactly the sum of the exchanged volume on the market. But we keep these things in mind for a future article with a different price formulation.
\end{itemize}

The model defined by the formulas  (\ref{intenuni}) and (\ref{priceuni}) corresponds to a univariate stochastic process. If one tries to incorporate some dependence between the assets prices, and to  have for example the Epps effect, and/or the lead-lag effect, one a priori needs  to extend the model to a  a multivariate form. The question is  where  we have to assume $\nu_{ij}\neq0$ and $h_{i}\neq0$ between the two assets? Of course we could introduce dependency between each component, but it is well known fact in financial modeling that a very complicated model will be very difficult to calibrate, instable, serve mainly for academics and not for practitioners. Then, we assume only two types of interaction: the  dependence for upward jumps in ask side in both assets, and the same for the  downward jumps. Since all  coefficient of the branching  matrix are assumed to be positive or equal to zero, if the two assets are negatively correlated, we don't have any other choice that to  introduce interaction between upward jump of $i$ and downward jumps of $j$. As in the univariate case\color{black}, we don't assume any interaction between upward and downward jumps  for the same asset so we delete one sign for the sake of simplicity.

Let us mathematically express the multivariate model. Writing $\lambda^{(i)}_{a,+}(t)$ for the intensity  of the  asset   $i=1,\cdots,d$ in the upward jump case and $\lambda_j(t)$ as in (\ref{intenuni}), asset $i$ is characterized by,

\begin{equation}
\begin{aligned}
&\text{Ask}^{(i)}:\left\{\begin{array}{rll}
\lambda^{(i)}_{a, +}(t)\!\!\!\!&=\lambda_{a, +}(t)\!\!\!\!&+\displaystyle\sum_{j\neq i =1}^d\nu^{(j)}_{a,+}\int_{(-\infty, t)\times \mathbb{R}} h_{a,+}(t-s)g^{(j)}_{a,+}(v)N^{(j)}_{a,+}(\mathrm{d}s\times\mathrm{d}v)\\
                                                 &\!\!\!\!&+\displaystyle\sum_{j\neq i =1}^d\nu^{(j)}_{a,-}\int_{(-\infty, t)\times \mathbb{R}} h_{a,-}(t-s)g^{(j)}_{a,-}(v)N^{(j)}_{a,-}(\mathrm{d}s\times\mathrm{d}v)\\
\lambda^{(i)}_{a, -}(t)\!\!\!\!&=\lambda_{a, -}(t)\!\!\!\!&+\displaystyle\sum_{j\neq i =1}^d\nu^{(j)}_{a,-}\int_{(-\infty, t)\times \mathbb{R}} h_{a,-}(t-s)g^{(j)}_{a,-}(v)N^{(j)}_{a,-}(\mathrm{d}s\times\mathrm{d}v)\\
                                                 &\!\!\!\!&+\displaystyle\sum_{j\neq i =1}^d\nu^{(j)}_{a,+}\int_{(-\infty, t)\times \mathbb{R}} h_{a,+}(t-s)g^{(j)}_{a,+}(v)N^{(j)}_{a,+}(\mathrm{d}s\times\mathrm{d}v)\end{array}\right.\\
&\text{Bid}^{(i)}:\left\{\begin{array}{rll}
\lambda^{(i)}_{b, +}(t)\!\!\!\!&=\lambda_{b, +}(t)\!\!\!\!&+\displaystyle\sum_{j\neq i =1}^d\nu^{(j)}_{b,+}\int_{(-\infty, t)\times \mathbb{R}} h_{b,+}(t-s)g^{(j)}_{b,+}(v)N^{(j)}_{b,+}(\mathrm{d}s\times\mathrm{d}v)\\
                                                 &\!\!\!\!&+\displaystyle\sum_{j\neq i =1}^d\nu^{(j)}_{b,-}\int_{(-\infty, t)\times \mathbb{R}} h_{b,-}(t-s)g^{(j)}_{b,+}(v)N^{(j)}_{b,-}(\mathrm{d}s\times\mathrm{d}v)\\
\lambda^{(i)}_{b, -}(t)\!\!\!\!&=\lambda_{b, -}(t)\!\!\!\!&+\displaystyle\sum_{j\neq i =1}^d\nu^{(j)}_{b,-}\int_{(-\infty, t)\times \mathbb{R}} h_{b,-}(t-s)g^{(j)}_{b,-}(v)N^{(j)}_{b,-}(\mathrm{d}s\times\mathrm{d}v)\\
                                                 &\!\!\!\!&+\displaystyle\sum_{j\neq i =1}^d\nu^{(j)}_{b,+}\int_{(-\infty, t)\times \mathbb{R}} h_{b,+}(t-s)g^{(j)}_{b,+}(v)N^{(j)}_{b,+}(\mathrm{d}s\times\mathrm{d}v).\end{array}\right.\\
\end{aligned}\label{intenmulti}
\end{equation}

 Hence, first part of any intensity correspond to the univariate case, second one is interaction in the case of positive dependance between assets and third one is the negative dependence between assets. For exemple, conditional intensity $\lambda^{(i)}_{a,+}$ is drawn by it's self exciting part, interaction with the upward jumps on best bid price and, interaction with upward and downward jumps on best ask prices of other assets. Of course, if the model is quite relevant, if asset $i$ is EUR/USD and $j$ is EUR/GBP, or USD/JPY and GBP/JPY (JPY = Japonese Yen), the branching coefficient for upward jumps in the best ask price should be positive while it should be close to zero for downward jumps like these two assets seem to fluctuate more or less together the last year. Conversaly, we should have a coefficient close to zero for upward jumps between the EUR/USD and the USD/CHF (CHF =Swiss franc) and positive for downward jumps. \color{black} To be  clear, we present in Table (\ref{intertable}), the interactions that we take into account:

\begin{table}[h!]\footnotesize
\begin{tabular}{c||cccc|cccc}
                 & ask$^{(i)}$ +  & bid$^{(i)}$ +  & ask$^{(i)}$ -    & bid$^{(i)}$ -    & ask$^{(j)}$ + & bid$^{(j)}$ + & ask$^{(j)}$ - & bid$^{(j)}$ - \\\hline\hline
ask$^{(i)}$ +    &   x            &   x            &    -             &    -             &   x            &   -            &    x             &    -   \\
bid$^{(i)}$ +    &   x            &   x            &    -             &    -             &   -            &   x            &    -             &    x   \\
ask$^{(i)}$ -    &   -            &   -            &    x             &    x             &   x            &   -            &    x             &    -   \\
bid$^{(i)}$ -    &   -            &   -            &    x             &    x             &   -            &   x            &    -             &    x   \\\hline
ask$^{(j)}$ +    &   x            &   -            &    x             &    -             &   x            &   x            &    -             &    -   \\
bid$^{(j)}$ +    &   -            &   x            &    -             &    x             &   x            &   x            &    -             &    -   \\
ask$^{(j)}$ -    &   x            &   -            &    x             &    -             &   -            &   -            &    x             &    x   \\
bid$^{(j)}$ -    &   -            &   x            &    -             &    x             &   -            &   -            &    x             &    x   \\
\end{tabular}
\caption{'x' if true, '-' if false. $i\neq j = 1,\cdots,d$. Notice is nothing else than the branching matrix $\boldsymbol{\nu}$. As we can see, in the case of two assets, we have for each quantity, four components, it self to produce the auto-excitation effect and two other, the corresponding quantities in the other asset, and bid with similar jump on the two assets. Up left and bottom right quart correspond to monovariate case $i$ and $j$.}
\label{intertable}
\end{table}
\normalsize

\section{Estimation Procedures}

Let us explain our estimation technique. There are several approaches to estimate the parameters in our model.  To simulate a  point process, we can apply thinning algorithm in \cite{DaVe03}, \cite{Li09} or \cite{Og81}. Of course, before applying one of them, we need to estimate the parameters of the model. The most natural estimation is given by the maximum of likelihood method . Let $\Theta$ be the parameters set which depend on mark distribution $\bold{f}=(f_1,\cdots,f_d)$. also consider the impact function $\bold{g}=(g_1,\cdots,g_d)$, the decay kernel $\bold{h}=(h_1,\cdots,h_d)$, the branching matrix, $\boldsymbol{\nu}=\{\nu_{ij}\}_{i,j=1,\cdots,d}$ and $\boldsymbol{\mu}=(\mu_1,\cdots,\mu_d)$. Let $I=[T_-,T^+]$ be  interval containing all arrival times. With these notation, the likelihood function is given by:

\begin{equation}
L(\{t_i,v_i\};\Theta) = \prod_{j=1}^d \int_{I\times\mathbb{R}} \lambda_j(t,v(t)|\mathcal{F}_{t})N_j(\mathrm{d}t\times\mathrm{d}v)\exp(-\Lambda_j(T^+)),
\end{equation}
where $\Lambda_j(T)$ is the compensator, or integrated intensity given by

\begin{equation}
\Lambda_j(T)=\int_{-\infty}^{T}\lambda_j(t,v|\mathcal{F}_t)\mathrm{d}t\times\mathrm{d}v,\:\: j\in\{1,\cdots,d\}.
\end{equation}

The discrete version of log-likelihood is then expressed as follows:

\begin{equation}
\log L(\{t_i,v_i\};\Theta) =  \sum_{j=1}^d\sum_{k=1}^N \log\lambda_j(t_k,v(t_k)|\mathcal{F}_{t_k}) -\sum_{j=1}^d\Lambda_j(T^+). \label{loglike}
\end{equation}
The maximization of log-likelihood could be done using a classical optimization algorithm. Notice that Bacry et al. \cite{Ba11a} proposed an alternative approach to estimate the parameters: the first find empirically the parameter set  $\Theta$ and the one minimizes the mean square error between the empirical signature plot and its estimate.\\

As we have said before, mark impact functions has to be normalized, and in order to do this we have to know their distribution function. To perform our analysis, we use two parities, EUR/USD and EUR/GBP recorded in milliseconds from January 30, 2012 00:00:00 to  March 09, 2012 21:59:00, one month ans one week. During this period, we have recorded 3 352 809 trades in EUR/USD and 2 178 009 for EUR/GBP. As the number of data is too large, we cut the historic week and averaged each result, then, we have 5 week for each parity. Data were cleaned using classical procedure, see \cite{Da01}. Hence, we plot cumulative distribution $\mathbb{P}(V>x)$ of ask volume which has produces an upward and downward jump separately at corresponding time, idem for bid volume. Results are given on figure (\ref{distribvolplot}). In order to find the mark distribution and therefore to compute likelihood function, two fits are proposed. \\

First fit are classically obtained with Gaussian law, $f_G(x)=\frac{1}{\sigma\sqrt{2\pi}}\exp(\frac{-(x-\mu)^2}{2\sigma^2})$, where $\mu$ is the mean and $\sigma$ the standard deviation. The second one is an exponential law, $f_E(x)\propto \exp(-\beta x), \beta>0$. \\

Not surprisingly, the Gaussian fit gives the worst fitting, tails are not described correctly, they vanish too quickly. The exponential law provides a very good fit of the bid and ask volume distribution. What is very interesting to see on this plot, on EUR/USD as on EUR/GBP, is that the cumulative distribution for ask volume when occurs an upward jump was very close to bid volume when occurs a downward jump. We have the same characteristic for the ask volume when occurs a downward jump and for the bid volume when occurs an upward jump. There are two kinds of orders that move the price, namely the market orders who take all available shares and limited orders between the quotes.\\

When a limited order to buy some shares is passed inside the quotes, it increases the bid price (upward jump on bid side). Reciprocally when an limited order appears bellow the ask price, it decreases ask price (downward jump on ask side).

When a buy marked order take all available shares, ask price increase to the next limit available in the limited order book, (upward jump on ask side).Reciprocally, when a sell market order take all available shares of the first limit, bid price below to the next limit available on the order book (downward jump on bid side).\\

We clearly see on figure (\ref{distribvolplot}) the two behaviors distinctly. Quantity offer by a new limited order inside the quotes was greater than the quantity offered by an 'old' limited order which becomes the new bid or ask price.\\

\begin{figure}[h!]
\begin{minipage}[b]{0.5\linewidth}
       \centering \includegraphics[height=7.5cm, width=6.5cm]{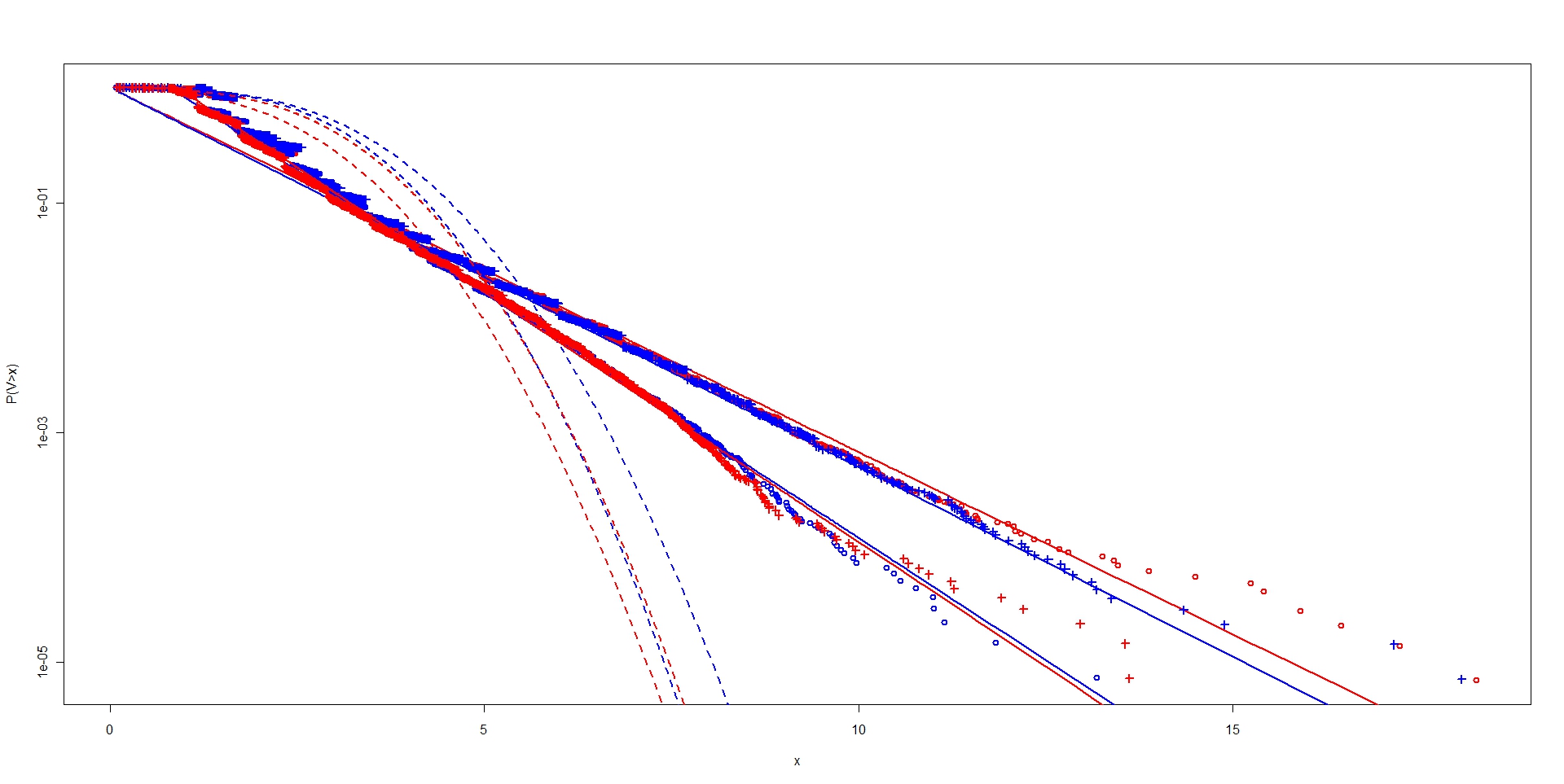}
\end{minipage}\hfill
\begin{minipage}[b]{0.5\linewidth}
       \centering \includegraphics[height=7.5cm, width=6.5cm]{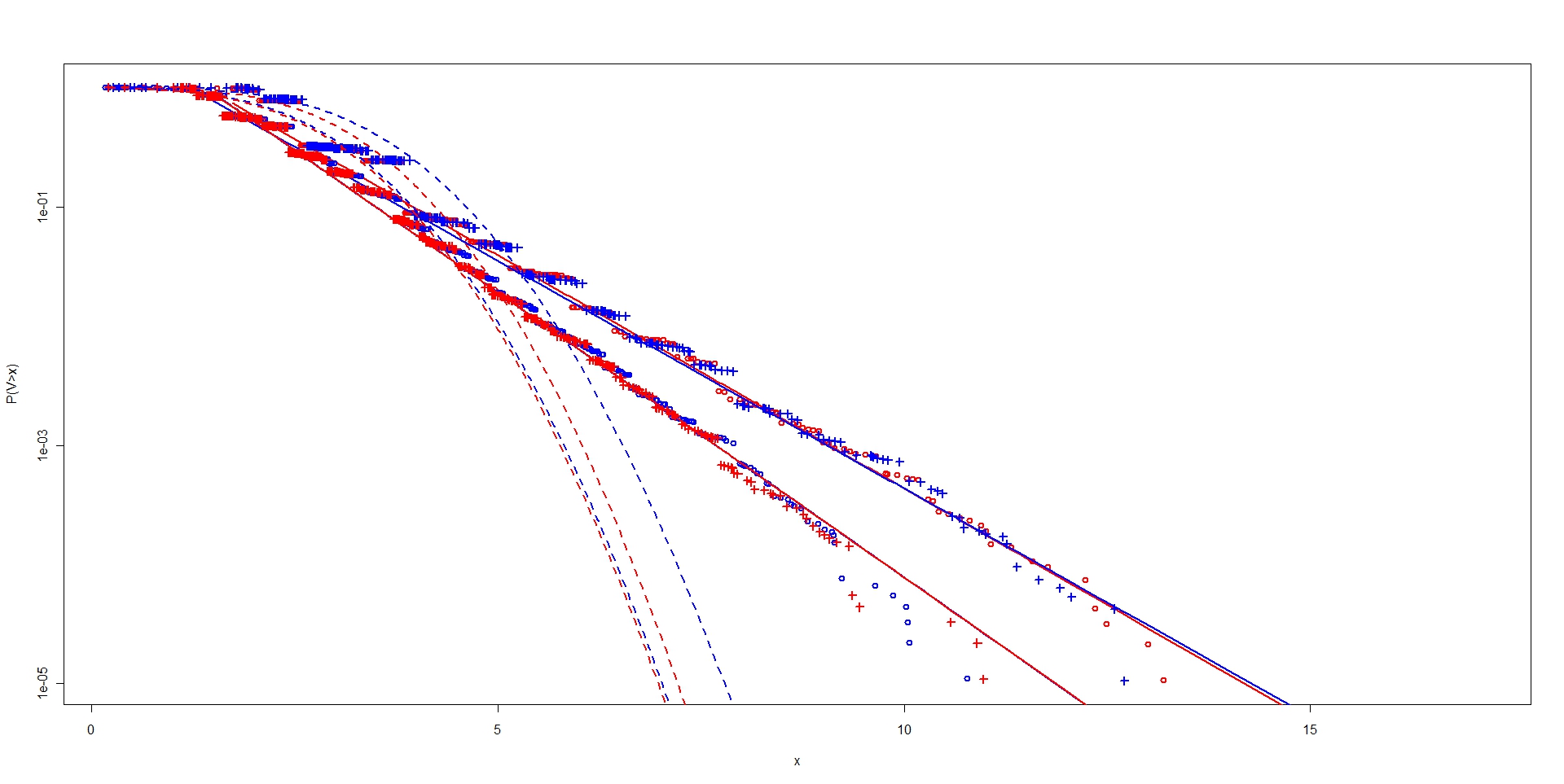}
\end{minipage}
\caption{Cumulative distribution in log scale of ask volume in a case of upward jump (blue circle), ask volume in a case of downward jump (blue cross), bid volume in a case of upward jump (red circle), bid volume in a case of downward jump (red cross) on EUR/USD (left) and EUR/GBP (right), from January 30, 2012 00:00:00 to March 09, 2012 21:59:00. Dashed line, gaussian fit, solid line, exponential fit.}
\label{distribvolplot}
\end{figure}

Hence, we will choose exponential law $\beta\exp(-\beta x)$, $\beta>0$. A suitable impact function is the power law, $x^\alpha$, take into account normalizing condition we have:

\begin{equation}
g(x)=\frac{\beta^\alpha}{\Gamma(\alpha+1)}x^{\alpha}.
\end{equation}

 Since the conditional intensity is the probability to observe a new event at time $t+\delta t$, the impact function quantifies the impact of the volume on the trade frequency. Thus, knowing empirical result, e.g. \cite{LiFaMa03}, we see that the exchanged shares with respect to price fluctuation seem envolve according a power law. Like this the instant of trade who produce price fluctuation, we choose a power law impact function. As we have explained before, market impact is the shifting induce by a trade. Simple example to price impact is to want to buy $x$ shares of an asset at market price, but, if available volume at market price is strictly less than $x$, say $x_m$, agent will buy $x_m$ shares at market price and $x-x_m$ at the next limit, and so one if the available share at next limit is strictly less than $x-x_m$. Thus, a natural problem is trading strategy for liquidition of large volume, we can cite for example the seminal work of R. Almgren, see e.g. \cite{Al09}, where the impact function is a power law. \\

We still have an important number of parameters to estimate. We have sufficient information to perform maximization of (\ref{loglike}) with classical optimization algorithm. Looking in detail theory of marked Hawkes process give us some way to simplify calibration.

The decay kernel $h_{i}(t-s)$ gives us the probability that an event of type $i$ will trigger an event of type $j$ at time $t$. In other words, it is the distribution of waiting time $t-s$ between the impact of event $i$ to the system at time $s$ and the occurrence of an event $j$ at time $t$. Brunching parameters $\nu_{ij}$ is the average number of event $j$ triggered by event of type $j$. This two quantities can be estimates with usual statistical techniques.\\

 Finally, we do not have assume any restriction on the value of the parameters. As we have say before, we admit an interaction between upward jumps on best bid and best ask price, and similarly for downward jump. Since the price process  is of course drawn by a certain probability, it is possible to observe after the simulations that the  ask and bid prices diverge. Hence, to avoid this problem we assume that instantaneous intensity for upward jumps on the ask and the instantaneous intensity for upward jumps are equals, $\mu^{(i)}_{a+}=\mu^{(i)}_{b+}$. The same assumption is considered  for the downward jumps $\mu^{(i)}_{a-}=\mu^{(i)}_{b-}$.

 After the estimation procedure for the parities, EUR/USD and EUR/GBP, we have found a spectral radius between 0.71 and 0.84 for each week and its averages is  0.81. Hence, the Hawkes process is well defined.

We now present the  conditional marked intensities realization corresponding to our ask and bid model on EUR/USD on figure (\ref{askinten}) for ask prices and (\ref{bidinten}) for bid prices. For a plot more comprehensive, we have plotted 2000 dots in 'deci'-seconds, then, a total of 3 minutes and 23 seconds, form 30-01-2012, 00:12:53 to 00:16:32. The procedure using  data on deci-seconds is the classical interpolation formula known as the tick estimator:
\begin{equation}
t=\text{arg }\max_{t_i}\{t_i\leq t\}, i=1,2,\cdots
\end{equation}
on the regular grid $\Lambda_t=\{0$sec.10, 0sec.20, $\cdots\}$ (see \cite{Da01} for more details).
 As we can see on the plot, it is possible to observe variation of the intensities while the corresponding quantities do not vary. In fact, since we have assumed interactions between upward jumps on ask price and upward jump on bid price, if the bid price increase, the conditional intensity $\lambda_{a+}$ will increase whereas ask price don't have increase.

\begin{figure}[h!]
       \centering \includegraphics[height=7cm, width=13cm]{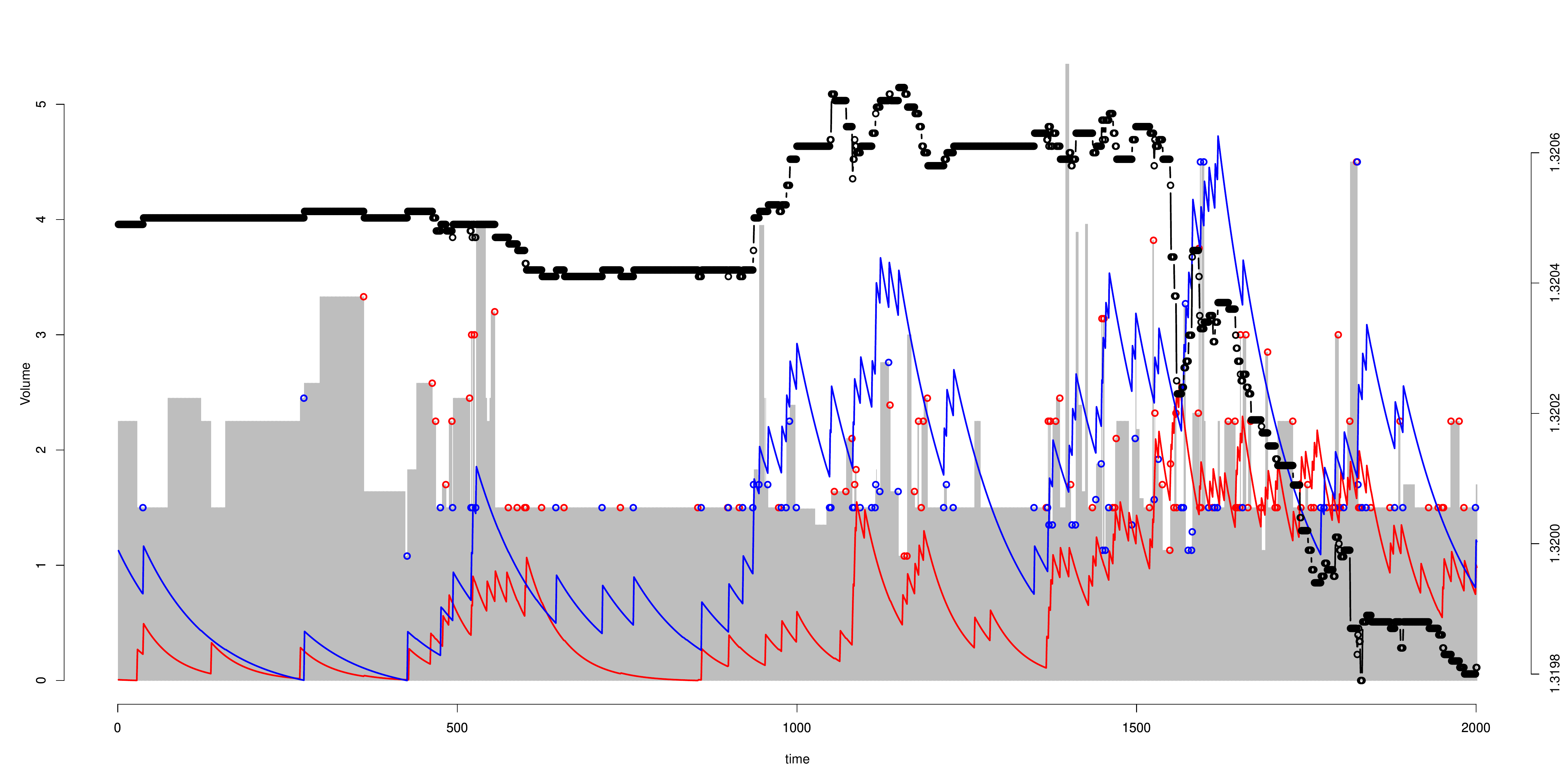}
\caption{Ask volume in gray, blue (red) circle correspond to upward (downward) jumps corresponding conditional marked intensities are plotted in line blue and red. Black dots are ask prices.}
\label{askinten}
\end{figure}

\begin{figure}[h!]
       \centering \includegraphics[height=7cm, width=13cm]{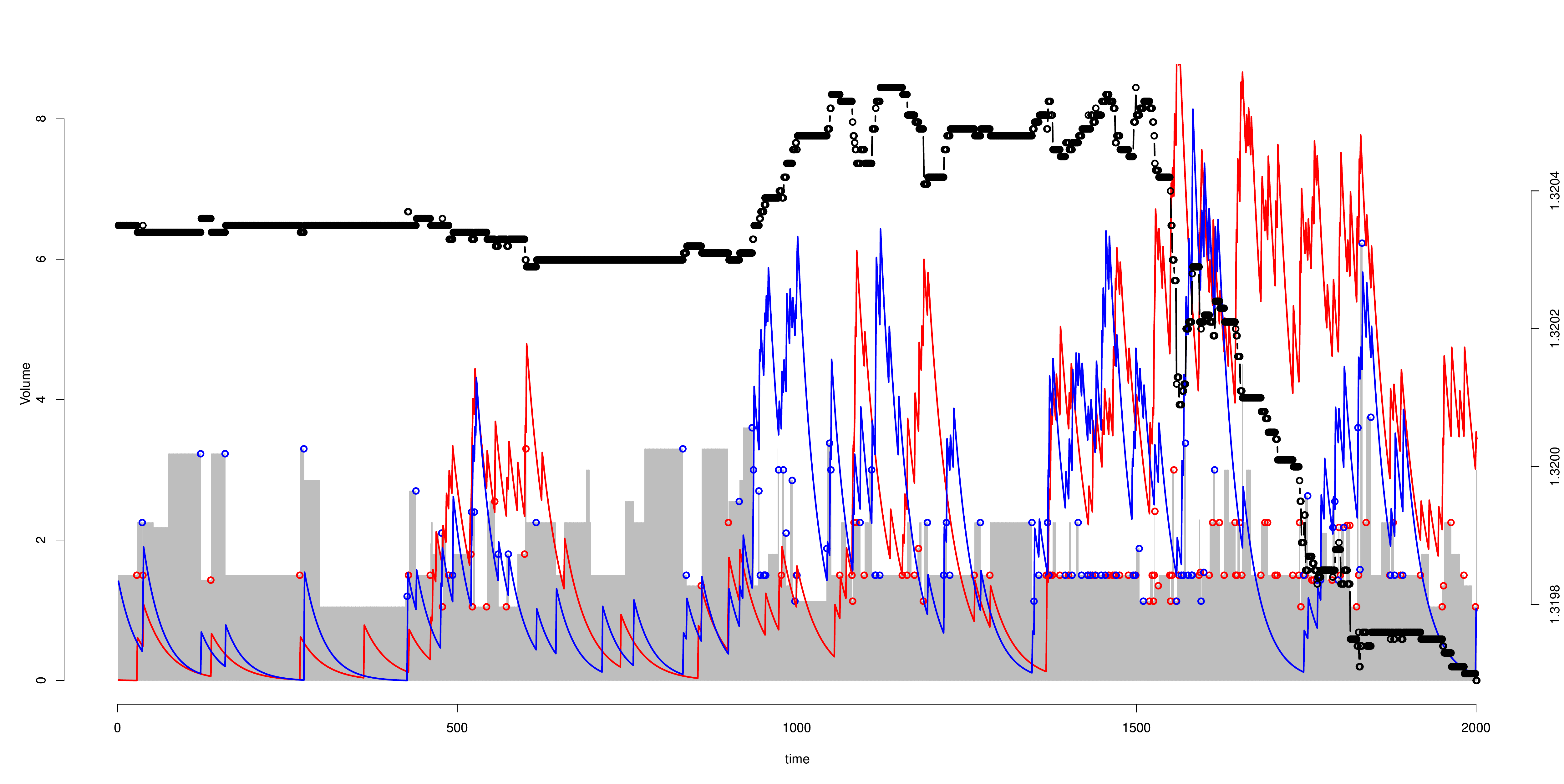}
\caption{Bid volume in gray, blue (red) circle correspond to upward (downward) jumps corresponding conditional marked intensities are plotted in line blue and red. Black dots are bid prices.}
\label{bidinten}
\end{figure}

\begin{figure}[h!]
       \centering \includegraphics[height=9cm, width=14cm]{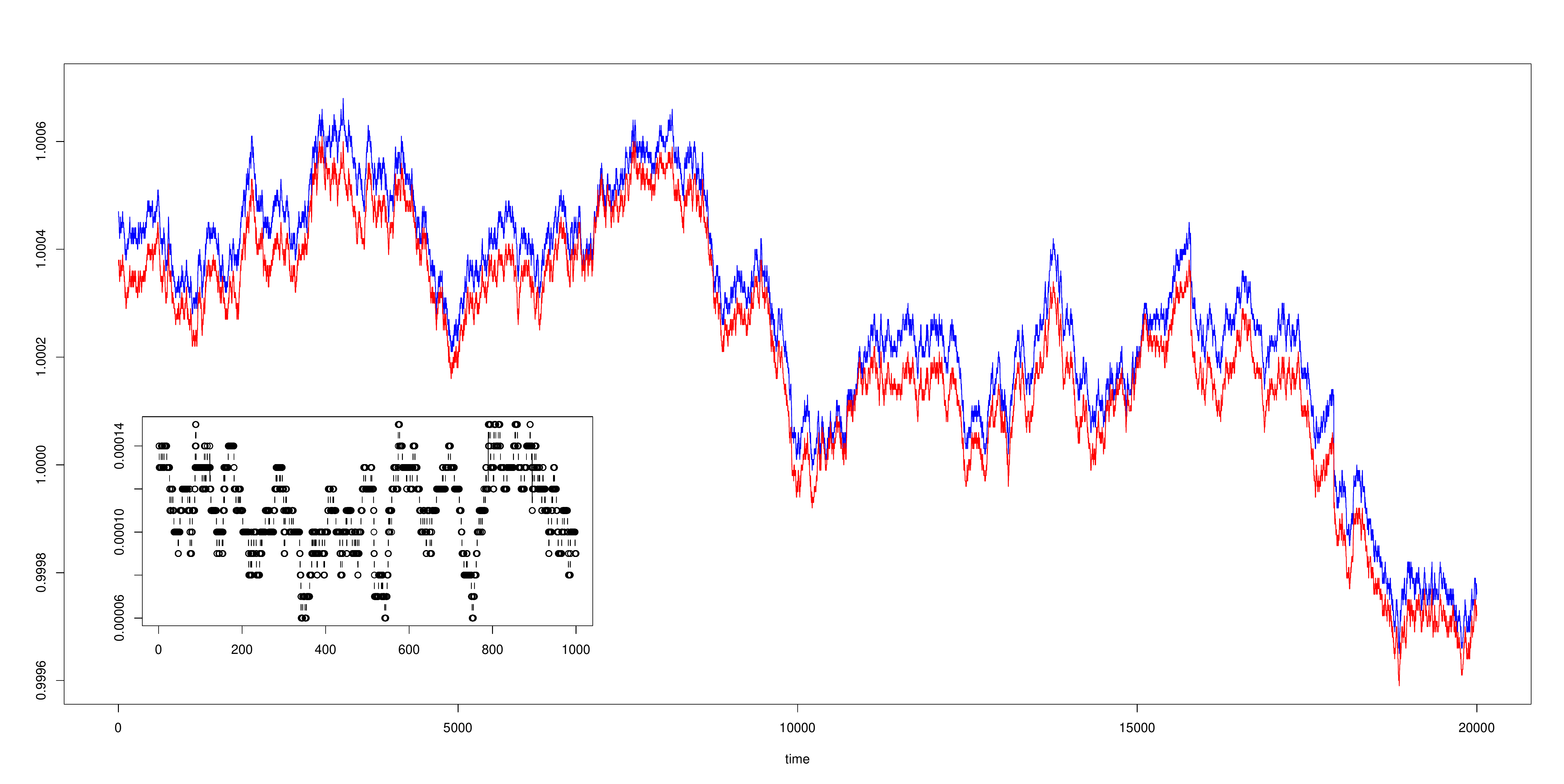}
\caption{Blue: best ask price simulation, red: best bid price simulation. Inset: corresponding bid-ask spread.}
\label{}
\end{figure}

\section{Stylized Fact}

Our study is focused on Forex markets, and in this situation there is not a "last price", there is  only bid and  ask price. The Forex market is what is usually  called an "OTC" (Over the Counter) market. In contrast to other markets, such as options or futures markets, there is no central reporting facility for OTC markets (in particular for Forex). The Forex market is international in the sense that each market place takes the relay of the other, trades are continuously recorded, they are no closing hours such as  in the stock market, hence parities evolve without interruption throughout  the whole day. It opens with the Australian (Sydney) market at 20:15 GMT on Sunday and it closes with the US (New-York) market at 22:00 GMT Friday. Hence  there is no official "tape".  The last traded price for Forex entirely depends  on  where you look. Bloomberg, Reuters, Yahoo, Google, IB, etc. have  each one different combinations of pools of liquidity from which they are gleaning this information. Among the major interbank market, we have Deutsche Bank, City and Barclays Investissement Bank.\\

If one looks at the financial time series with sufficiently low frequency, for example in daily or weekly scaling, even little higher if the considered asset is very liquid, for example hourly, we observe a continuous process. For higher frequency, the continuous time hypothesis is clearly rejected. The price of an asset, currency, stock, commodity and so on, is the consequence of a the  transactions which have been made. If all agents buy an asset, the price will tend to increase, inversely, if all agents sell an asset, the price will tend to decrease. It  is the interaction between agents which creates the price fluctuation that  we observe. This processes is not continuous in time, this is not necessarily at each milli-second that an agent finds a counterpart to buy (resp. sell) what he wants to sell (resp. buy), hence, the price stops moving until deal happen. That's why fluctuation has to be model as a point process.\\

As told before, there are some characteristics of very high frequency financial data. For each study of these quantities, we will compare the empirical result, the best fit obtained with an appropriate law and the realization given by the model.\\

Let $X(t)=\log p(t)$, the log price of an asset. The signature plot, introduced in \cite{An00}, corresponds to the  volatility evolution on $[0,T]$ with respect to frequency,

\begin{equation}
\mathbb{V}_X(\tau)=\frac{1}{T}\sum_{n=0}^{\lfloor T/\tau \rfloor }(X((n+1)\tau)-X(n\tau))^2\label{sig}
\end{equation}
 so more the lag $\tau$ is large,  more this quantity is decreasing in power law. The explanation of that signature is quite simple: at very high frequency, data arrives at more or less distant time, asset prices are discontinuous and show jump, hence, volatility tend to increase with the frequency because of these jumps. Conversely, the fluctuations observed at large scale tend to smooth evolutions, hence, volatility tends to decrease until their equilibrium value. This effect is plotted on (\ref{sigplot}) using mid price quote by averaging week by week.\\


\begin{figure}[h!]
       \centering \includegraphics[height=6cm, width=13cm]{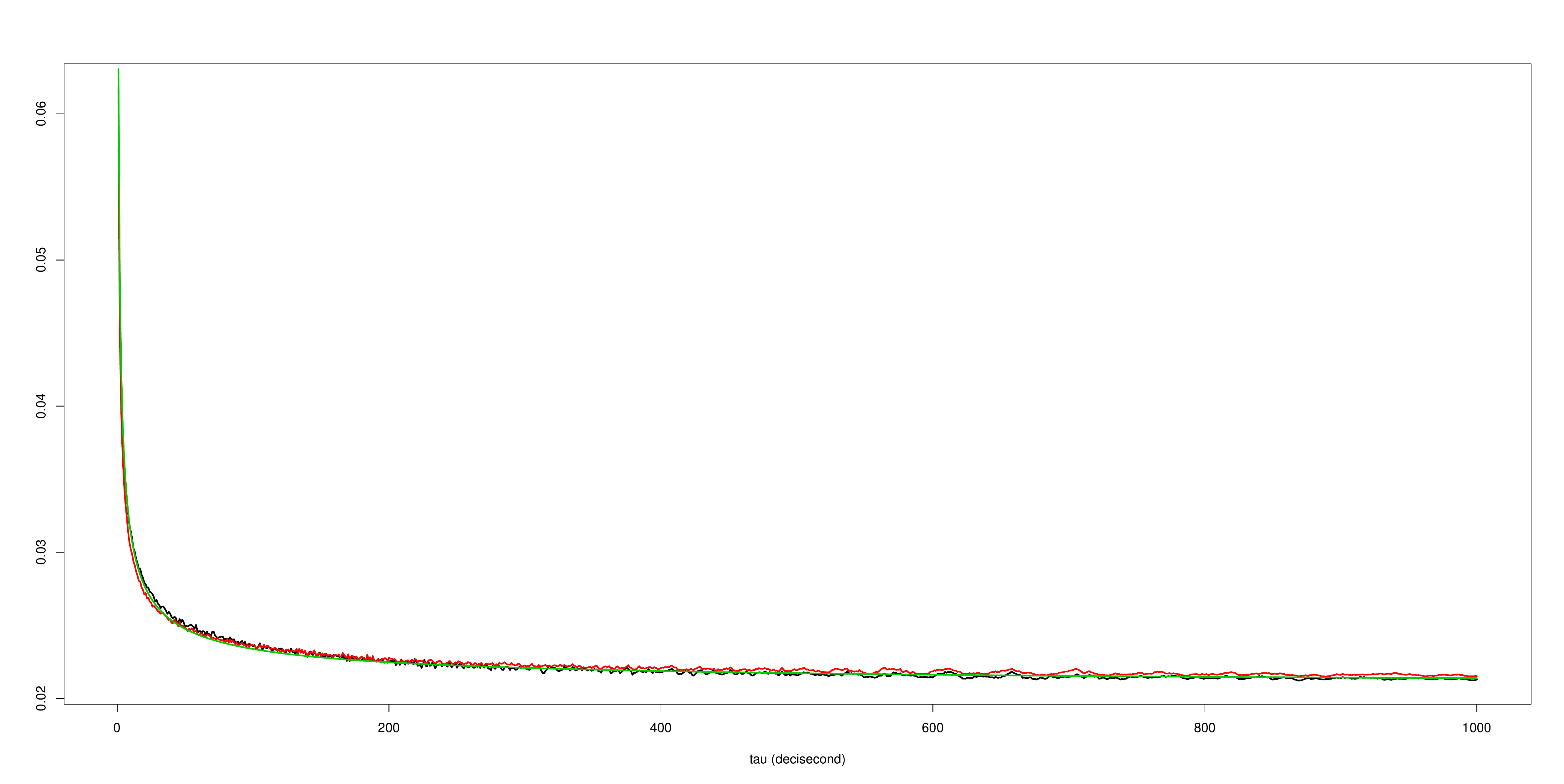}
\caption{Signature plot of EUR/USD in black and in red for the simulation, fit in power law (green), $\tau$ varying for 0 to 100 seconds.}
\label{sigplot}
\end{figure}

The Epps effect, introduced in \cite{Ep79}, is the evolution of the correlation between two financial assets with respect to the frequency. As frequency increases, the correlation tends to vanish. This not very complicated to realize, if the  orders arrive at discontinuous time. On the other hand this is not easy to observe if the orders, say sell orders, of the two considered financial assets arrive at exactly the same times and then, we observe a correlation, although if we are looking for two assets very liquid, it  will be always a lag and the correlation tends to vanish at very high frequency.

\begin{equation}
\rho_{1,2}(\tau)=\frac{\mathbb{C}\text{o}_{1,2}(\tau)}{\sqrt{\mathbb{V}_{X_1}(\tau)\mathbb{V}_{X_2}(\tau)}}
\end{equation}
where $\mathbb{V}_X(\tau)$ is define like in (\ref{sig}) and the empirical covariation $\mathbb{C}\text{o}$ is given by,

$$
\mathbb{C}\text{o}_{1,2}(\tau)=\frac{1}{T}\sum_{n=0}^{\lfloor T/\tau\rfloor}\left(X_1((n+1)\tau)-X_1(n\tau)\right)\left(X_2((n+1)\tau)-X_2(n\tau)\right).
$$

This effect is plotted on (\ref{eppseffect}) using mid price quote by averaging week by week.

\begin{figure}[h!]
       \centering \includegraphics[height=6cm, width=13cm]{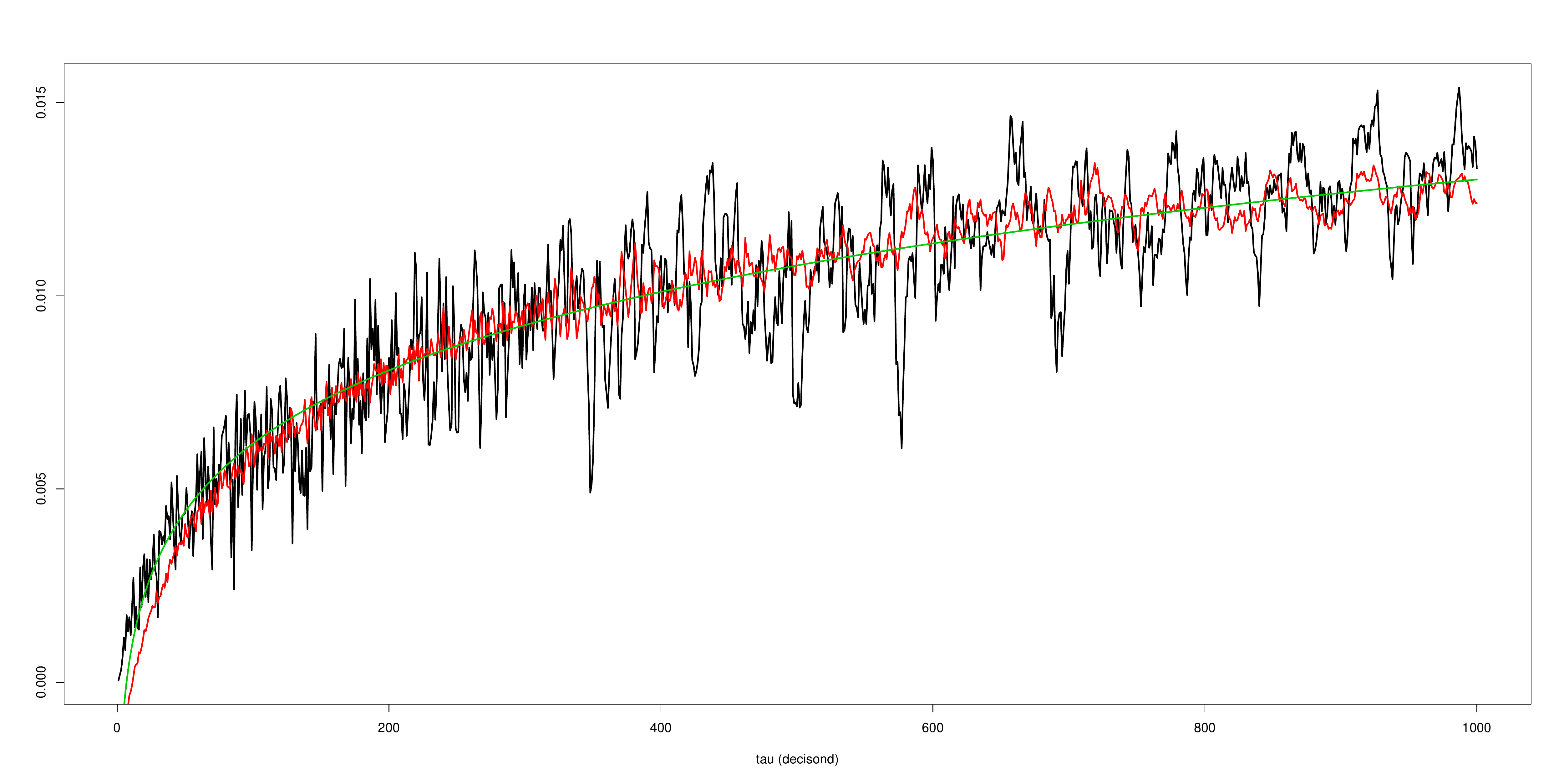}
\caption{Epps effect on the two considered currency foreign exchange data, EUR/USD and EUR/GBP, in black and in red for the simulation, corresponding fit in power law, $\tau$ varying for 0 to 100 seconds.}
\label{eppseffect}
\end{figure}

\section{Forecast and Application to Risk Management}

Even if we have performed our analysis on the particular case of the  Forex market, we can easily apply our model to various types of financial assets. The analysis of the  bid/ask price fluctuations gives us the spread, and then, an important fees component. We therefore can  deduce a very interesting stress test type for high frequency  trading strategies.

A well-known fact to practitioners is that, sometimes  it is not very complicated to find some "rules" in order to deduce the  next market move, but if one includes the  broker fees, market impact and bid-ask spread, the detected gains probabilities will reduce to zero, and there is  a cost to be  paid.

 Let us explain in details these kinds of transaction costs. The broker fees depend mainly on the broker and correspond in general to a percentage of effective value held.

The market impact is the price to be paid to pass an order. Since the available  volume on each limit is not infinite, if the amount of the desired volume is strictly bigger  than the available volume at the market price, the agent will take all shares at the  market price and then, he will  take shares to the next limit and so one, until he/she bought (sold) the desired  quantity. Hence, his transaction is equal to $V=\sum_{k_i}(p_a+x_ip_{\text{tick}})$ where $p_{\text{tick}}$ is the tick size, $x_i$ measures the distance between the best price and the next limit with $x_1=0$, $x_i\in\mathbb{N}^*$ and $k=\sum_{i=1}^nk_i$ is the quantity that he/she has buy at limit $p_a+x_ip_{\text{tick}}$. Finally, his fees  due to the  market impact is equal to $\sum_{i=1}^nk_ix_ip_{tick}$.

Consequently, the bid-ask spread is then the price to pay to an impatient trader who passes only orders at the  market prices. Since there is no a unique price on the market but a price for buy (ask) and a  price for sell (bid), a transaction consisting to buy at market price $k$ shares and closes its position at time $t+\delta t$ at the market price, will have a cost of $ks(t+\delta t)$, with $s$ denoting the spread.\\

To test high frequency trading strategies, incorporating the spread in the model is thus fundamental. The spread is directly deduced as  the difference of the two simulations of the bid and ask prices. To incorporate the  market impact, it  is quite more difficult. We need to know the next limit and the corresponding volume. Without loss of generality, we could argue that the next price after the best ask price, $p_a$, is equal to $p_a+p_{\text{tick}}$ and so one. Similarly for the bid price. This hypothesis is completely possible in a sufficiently liquid market, and  using the same notation as before, the market impact cost will then be  equal to $\sum_{i=1}^nk_ix_ip_{tick}=\sum_{i=1}^nk_ip_{tick}$. So the knowledge of the best quotes allows  to deduce the next limits. But, we do not know the available share on each limit. For the simulation we have simply drawn the volume process with an exponential law. Of course, statistical properties of the  exchanged volume are very rich and need an extensive study to build a robust model, see e.g. \cite{GoPlGaSt00}.

Finally, by using  the structure of the intensity of the Hawkes process, one can  deduce a certain forecast strategy. It is shown in \cite{Ve95} that the probability of next time event $\tau^*$ that  appears after time $\tau$ is given by the formula

\begin{equation}
\mathbb{P}(\tau^*>\tau)=\exp\left(-\int_0^\tau\int_\mathbb{R}\lambda(s,v|\mathcal{F}_s)\mathrm{d}v\mathrm{d}s\right).
\end{equation}
We can find such $\tau$ that maximizes the probability for all the $4$ intensities in the model (\ref{intenuni}) or for the $d*4$ intensities in (\ref{intenmulti}). To perform the simulation, notice that we have to update at any time $t+\delta t$ until $t+\tau$ history $\mathcal{F}_t$.  With this information, we could deduce what will be the next event and take the corresponding position. In contrast to  the most part of the literature which takes into account only mid or last price, knowing the probability that a downward or upward jump will happen on the bid or ask side give us an idea on the  next price movement, and also  the expected spread. This  information indicates if the trade has a chance or not to be profitable. Of course, to deduce the size of the next  bid-ask spread, we need to have an idea about the  next  bid ask price value.

 Producing the  mark at any time $t+\delta$ until $t+\tau$ is also needed to perform our simulation. The mark value, $v(t+\delta t)$ can be generated from the density: $f(v)=\frac{\lambda(t+\delta t, v|\mathcal{F}_{t+\delta t})}{\int_\mathbb{R}\lambda(t+\delta t, v|\mathcal{F}_{t+\delta t})\mathrm{d}v}$. Adding the generated pair $(t+\delta t; v(t+\delta t))$ to the history permits to perform the exercise and moreover gives us all information needed to the next time price movement in a risk management framework: the spread and the market impact.

\section{Conclusion}

In this paper we  propose a model that describes the first line of an order book, which is represented by the best bid and best ask prices. The main line of our work  is  to say that the arrival of the orders does not depend only on the past history, but also on the volume. To this end, the idea is to use a marked point process instead of an unmarked one. Small quantity of shares does not affect the market in the same as huge quantities do, and then, the arrivals intensities changes. These two quantities are closely linked.

We have briefly discussed about the compound case, see (\ref{compound}). We emphasize  that it could be interesting to study further  this representation. Finally, it seems  possible to extend our model to an important part of modern finance theory such as computing Value at Risk or option pricing via a Monte Carlo scheme for very high frequency data.

\end{document}